\documentclass{PoS}
\pdfoutput=1
\usepackage{subfig}

\title{Improved Lattice Spectroscopy of Minimal Walking Technicolor}

\ShortTitle{Improved Lattice Spectroscopy of MWT}

\author{\speaker{Eoin Kerrane}, Luigi Del Debbio\\
	School of Physics \& Astronomy, University of Edinburgh, EH9 3JZ, Edinburgh, UK\\
        E-mail: \email{eoin.kerrane@ed.ac.uk, luigi.del.debbio@ed.ac.uk}}

\author{Claudio Pica\\
	CP$^\mathit{3}$-Origins \& IMADA, University of Southern Denmark, Odense, Denmark\\
        E-mail: \email{pica@cp3.sdu.dk}}

\author{Agostino Patella\\
	CERN, Physics Department, 1211 Geneva 23, Switzerland\\
        E-mail: \email{agostino.patella@cern.ch}}

\author{Antonio Rago\\
	Department of Physics, Bergische Universit\"at Wuppertal, D-42119 Wuppertal, Germany\\
        E-mail: \email{rago@physik.uni-wuppertal.de}}

\author{Biagio Lucini\\
	School of Physical Sciences, Swansea University, Singleton Park, Swansea SA2 8PP, UK\\
        E-mail: \email{b.lucini@swansea.ac.uk}}

\author{Francis Bursa\\
        Jesus College, University of Cambridge\\
        E-mail: \email{f.bursa@damtp.cam.ac.uk}}

\author{Thomas Pickup\\
        University of Oxford\\
        E-mail: \email{pickup@thphys.ox.ac.uk}}

\author{David Henty\\
	Edinburgh Parallel Computing Centre, University of Edinburgh, EH9 3JZ, Edinburgh, UK\\
	E-mail: \email{d.henty@epcc.ed.ac.uk}
}

\abstract{We present an improved study of spectroscopic observables in the $SU(2)$ Yang-Mills theory with two adjoint fermions. We make an improvement on the
precision of previous results which clarify the scale of finite volume effects present. This analysis adds to the evidence for near-conformal dynamics of this
theory, while indicating a preference for a low anomalous mass dimension of the massless theory.
\begin{flushright}
CERN-PH-TH/2010-259,
CP3-Origins-2010-49,
WUB/10-32, 
Edinburgh 2010/34
\end{flushright}
}

\FullConference{The XXVIII International Symposium on Lattice Field Theory, Lattice2010,\\
		June 14-19, 2010,\\
		Villasimius, Italy}

\begin{document}

\section{Minimal Walking Technicolor}
The mechanism of dynamical electro-weak symmetry breaking (often referred to as technicolor) \cite{Weinberg:1975gm} remains a possible explanation for the
breaking of the electro-weak symmetry observed in nature. Gauge theories which possess an approximate infra-red fixed point have been proposed as preferred
candidates for the technicolor sector in models of extended technicolor \cite{Appelquist:1986an} as it has been argued that they
would allow suppression of flavour changing neutral currents while permitting generation of large fermion masses. This is achieved through the non-trivial
renormalisation dynamics between the technicolor and extended technicolor scales. This property of the theory has been coined ``walking'' in reference to the
slow running of the coupling between the two scales \footnote{For a recent technicolor review see \cite{Hill:2002ap}.}.

We can hope to generate an infra-red fixed point in a gauge theory, while minimising contributions to electroweak precision constraints, by adding a low
number of fermion flavours in higher gauge representations to a gauge theory with a low number of colours. For these reasons the gauge theory theory with symmetry
group $SU(2)$ and two flavours of Dirac fermion in the adjoint representation has been termed minimal walking technicolor (MWT) \cite{Dietrich:2006cm}.

Conformal or near-conformal dynamics have not been conclusively identified in any theory as yet, although there are encouraging hints from a number of sources. Due to the non-perturbative nature of the problem, much of the investigation of this problem has arisen from lattice simulations, and this activity is growing. Some recent lattice studies of MWT \cite{Hietanen:2009az,Bursa:2009tj,Bursa:2009we} have
attempted to identify a near-conformal behaviour directly from the behaviour of the coupling and anomalous dimensions of the theory under renormalisation flow.
Others \cite{Catterall:2007yx,DelDebbio:2008zf,DelDebbio:2009fd,Hietanen:2009zz,DelDebbio:2010hx,DelDebbio:2010hu}, including this work, perform measurements of physical observables in the
theory and from their behaviour attempt to identify signals of near-conformal dynamics.

This study builds on previous work in \cite{DelDebbio:2008zf,DelDebbio:2010hu}. In particular we seek to support these results and establish their reliability
by investigating the effect of performing measurements and analysis using alternative methods. In addition, through this we expect to ascertain the scale of
systematic uncertainties present, which have as yet been largely unexplored. 

\section{Signals of Conformality}
\label{conformal}

The question of whether MWT in the chiral limit posesses an actual infra-red fixed point, i.e. it lies within the conformal window, or an approximate fixed
point, has not been conclusively answered. Nevertheless, it is clear that MWT with a non-zero fermion mass and defined in a finite volume, as simulated on the
lattice, cannot be conformal. If the chiral continuum theory posesses an infra-red fixed point, the lattice results will be described by a mass-deformed
conformal gauge theory. Recent discussions of scaling laws in such theories \cite{DelDebbio:2010hu,DelDebbio:2010jy,DelDebbio:2010ze} have derived scaling
relations for hadronic masses and amplitudes in terms of the mass deformation. 

In a theory with strong chiral symmetry breaking (like QCD), the mesonic spectrum departs significantly from degeneracy. The pseudoscalar mesons become massless
in the chiral limit, while the rest of the spectrum is expected to retain a finite mass.
In a mass deformed conformal theory, the outlook is different. In approaching a conformal limit, the theory respects the hyperscaling property, whereby
all masses $M$ in the theory scale identically. They must vanish in the limit of vanishing quark mass $m$. From an analysis of the
renormalisation of mesonic two-point functions we can deduce that all meson masses $M$ in the theory vanish as $M\sim m^\rho$ where the critical exponent $\rho$
is given by $\rho=\frac{1}{1+\gamma_\ast}$ where $\gamma_\ast$ denotes the anomalous dimension of the chiral condensate  in the conformal theory at the
infra-red fixed point. This quantity is phenomenologically very interesting as it determines the extent to which the chiral condensate is enhanced at the
extended technicolor scale with respect to the technicolor scale when a near-conformal gauge theory is included as the new strongly coupled gauge group in a
model of extended technicolor. We will attempt to fit our lattice data with this hyperscaling relation in order to determine whether the massless limit of MWT
is indeed a conformal theory, and also to deduce a range of $\gamma_\ast$ preferred by the data. 

In \cite{DelDebbio:2010hu} it is also shown that in a theory confined to a finite box of size $L$, analysing the renormalisation of the free energy density and
treating the box size $L$ as a relevant parameter, there exist universal finite-size scaling laws for physical observables. These can be summarised as
$LM\sim\Upsilon(Lm^\rho)$ for any observable $M$ of mass dimension one. The function $\Upsilon$ will differ for each $M$, but this relation allows us to compare
data across different lattice volumes in order to draw conclusions on the range of $\gamma_\ast$ which are preferred by the data.

\section{Systematic Spectroscopy}
\label{systematics}
This study builds on the work described in \cite{DelDebbio:2008zf,DelDebbio:2010hu} where spectroscopic observables of MWT were measured through lattice
simulations. The computation was performed using the \emph{HiRep} code, designed to simulate theories of general number of colours $N_c$ and number of flavours
$N_f$ of fermions in a generic representation $R$ of the gauge group. The simulations used the Wilson gauge action, and the Wilson fermion formulation along with the RHMC algorithm. A number of lattice volumes have been analysed,
from $16\times8^3$ to $64\times24^3$ with a range of bare quark masses. We describe analysis of ensembles at a single lattice spacing, with $\beta=2.25$.

For this study we have performed some alternative analyses to those in \cite{DelDebbio:2010hu}. We have modified the \emph{Chroma} suite of lattice software \cite{Edwards:2004sx} to
operate with $N_c\neq 3$ and a number of fermion representations $R$ other than the fundamental. Using the resulting code we have utilised the in-built smearing
routines found in \emph{Chroma} to perform measurements on the gauge configurations generated with \emph{HiRep} with a number of different quark smearings. 
 
The signals for the observables obtained from correlators using a wall-smeared source operator are the cleanest available. This could be expected due to the
enhancement of the overlap of the wall-smeared quark bilinear with the mesonic ground states due to its projection onto zero momentum. As a result it was
decided to produce a complete set of wall smeared correlators on our data, complementing the local correlators already analysed \cite{DelDebbio:2010hu}. We present full results for all
observables of interest in Sec. \ref{results}.

In addition to this investigation of the effect of smearing on the observables, we have also investigated possible systematic errors arising from our analysis
methods. Firstly, we considered the algorithm for extracting meson masses from the correlators. The Prony method \cite{DelDebbio:2010hu} using only the ground
state mass was preferred to the effective mass method \cite{DelDebbio:2007pz}, and so the Prony method was implemented in all analyses. We then examined whether
the effective observable for the PCAC quark mass and decay amplitudes was affected by the use of the effective meson mass or its fitted value. Finding little
variation, it was decided to use the effective mass in all definitions. 

\section{Results}
\label{results}

In the following we present results from the wall-smeared inversions on all available ensembles generated with HiRep. The analysis used every tenth
configuration in the Monte Carlo chain. Meson masses were extracted from the correlators using the Prony method using only the ground state mass. The definition
of effective observables for the PCAC mass and decay constants are those used in \cite{DelDebbio:2010hu} and can be found explicitly in \cite{DelDebbio:2007pz}.

\subsection{Systematics}

\begin{figure}[!htp]
\centering
\subfloat[Effective pseudoscalar masses on the $16\times8^3$ lattice.]{
\includegraphics[scale=0.21]{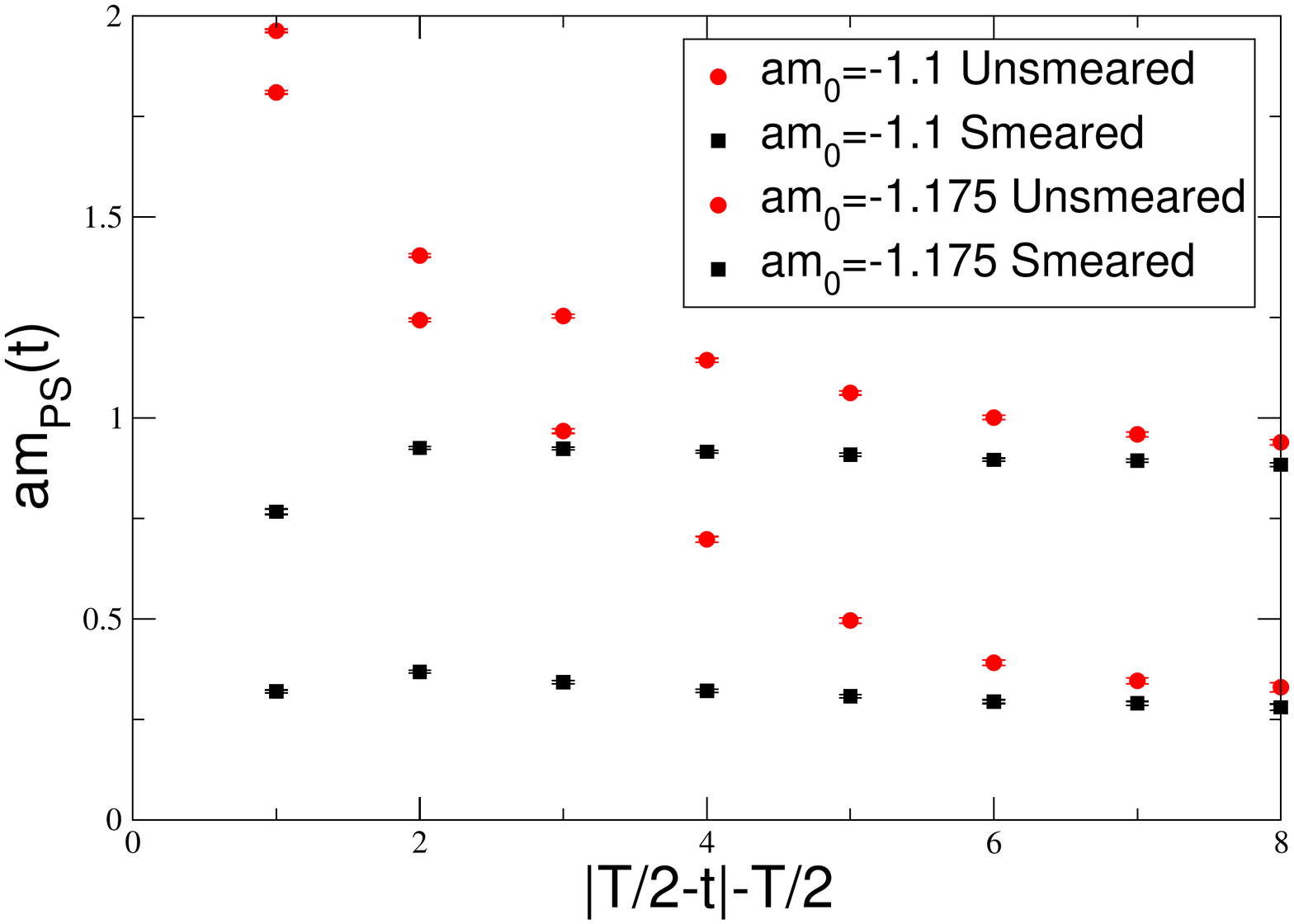}
\label{mpssys16}
}
\subfloat[Effective pseudoscalar masses on the $64\times24^3$ lattice.]{
\includegraphics[scale=0.21]{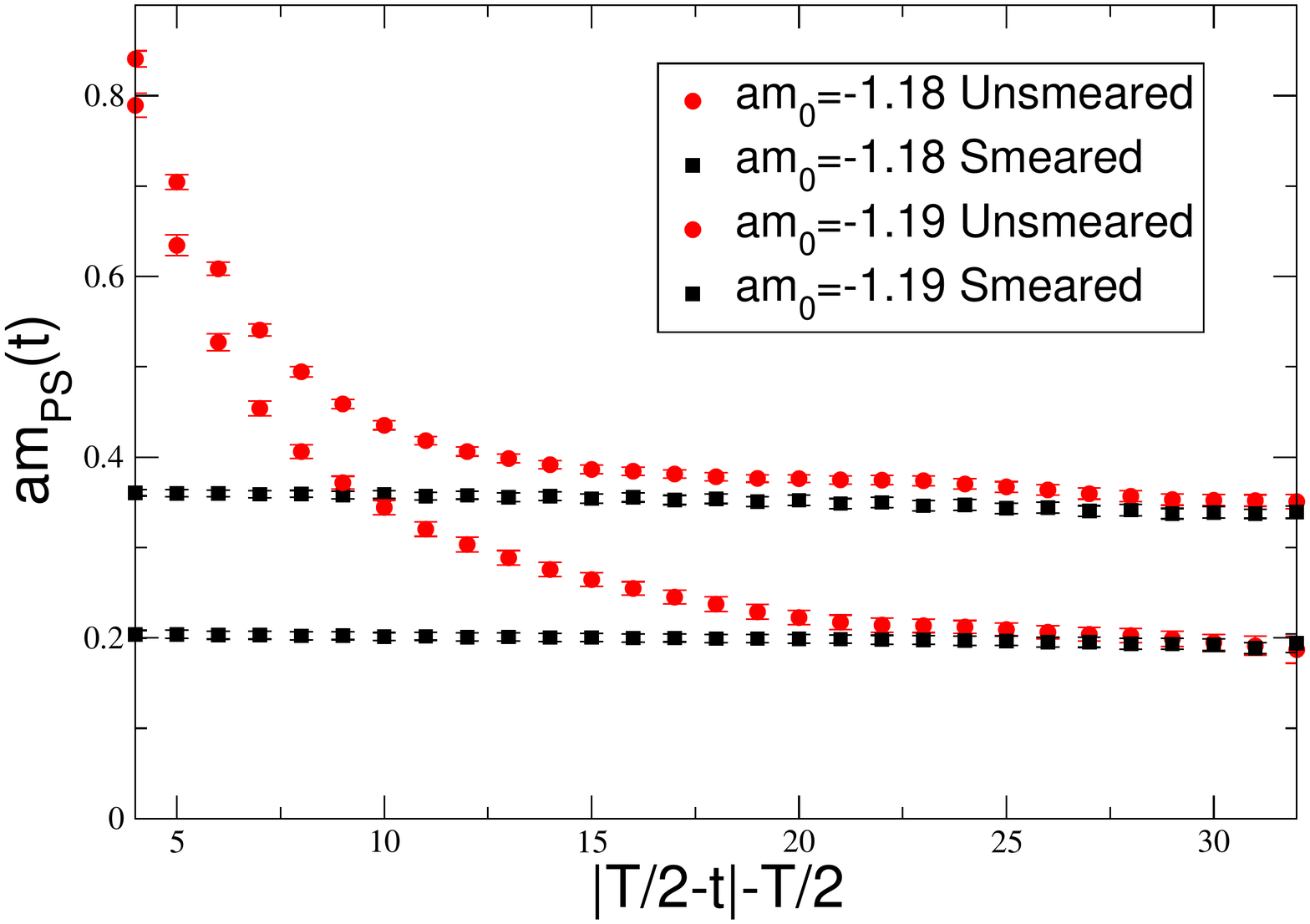}
\label{mpssys64}
}
\caption{Effective pseudoscalar masses from both smeared and unsmeared correlators.}
\label{mpssys}
\end{figure}

As stated in Sec. \ref{systematics} the wall-smeared correlators result in significantly improved signals for our mesonic observables.
This is illustrated in Fig. \ref{mpssys} for a choice of bare masses, on our
smallest and largest lattices. It can be seen that on the smallest lattice, the result from unsmeared correlators is contaminated with excited states and that their effects are not diluted even at the centre of the lattice. This will lead to a discrepancy between the results from the smeared
correlators and those from the unsmeared. In contrast however, we see that on the largest lattice, the temporal extent is sufficient to suppress the effect
of excited states in the local correlators towards the centre of the lattice, and the two results are in good agreement. As such we expect the smeared results
to systematically differ from the local results on smaller lattices while we expect agreement on the larger lattices. 
Below we present the smeared results, for the full local results see \cite{DelDebbio:2010hu}.

\subsection{PCAC mass}

In Fig. \ref{pcacres} results for the PCAC mass on all ensembles are presented. The inset illustrates a close up of the approach to the chiral limit, with
a linear extrapolation to zero quark mass. Using this we find the critical bare quark mass to be $am_c=-1.2025(5)$, 
which compares very well to the result obtained from the local data \cite{DelDebbio:2010hu}.

\subsection{Meson masses}

\begin{figure}[!htp]
\centering
\subfloat[PCAC quark mass.]{
\includegraphics[scale=0.21]{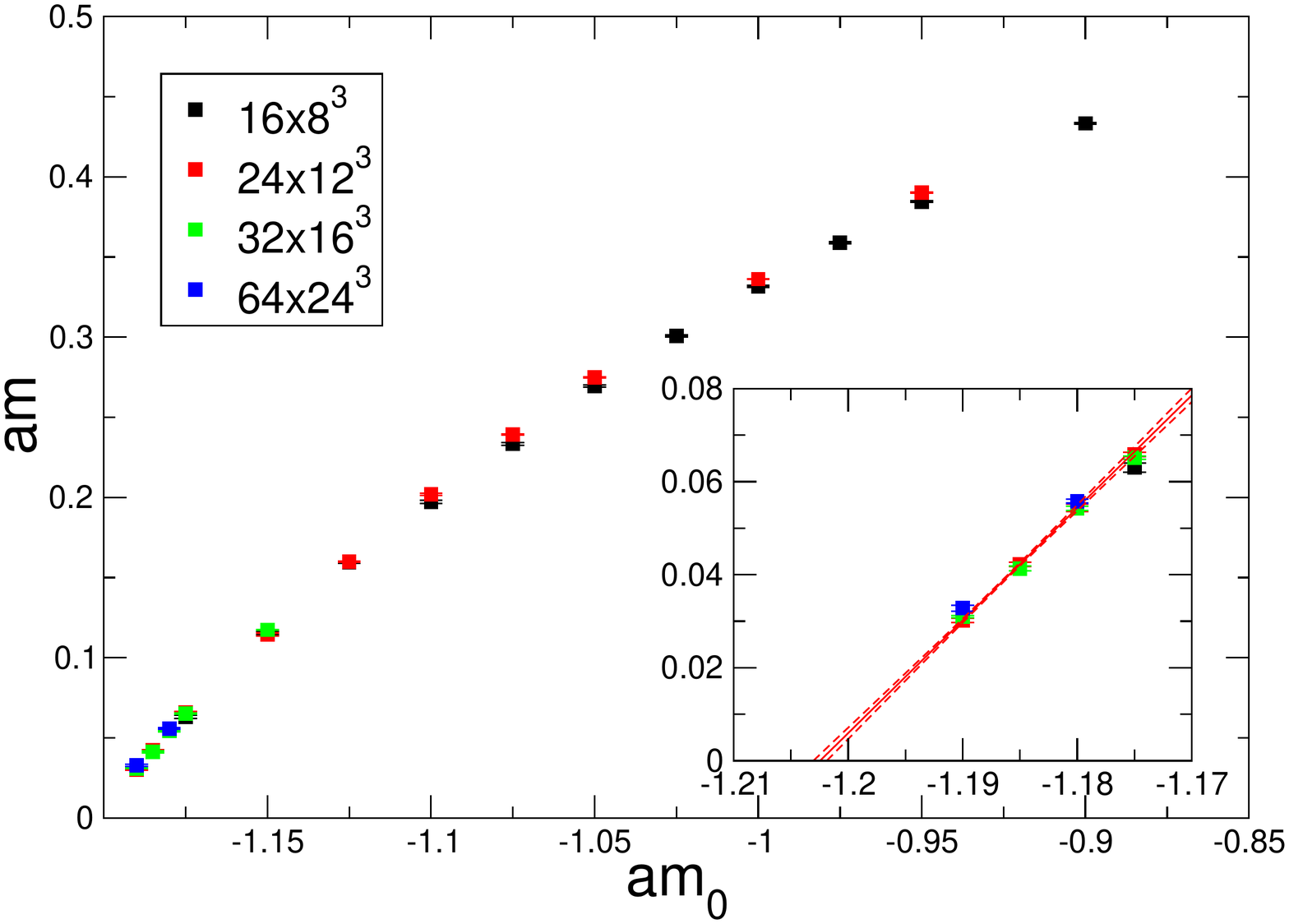}
\label{pcacres}
}
\subfloat[Pseudoscalar mass $m_{PS}$.]{
\includegraphics[scale=0.21]{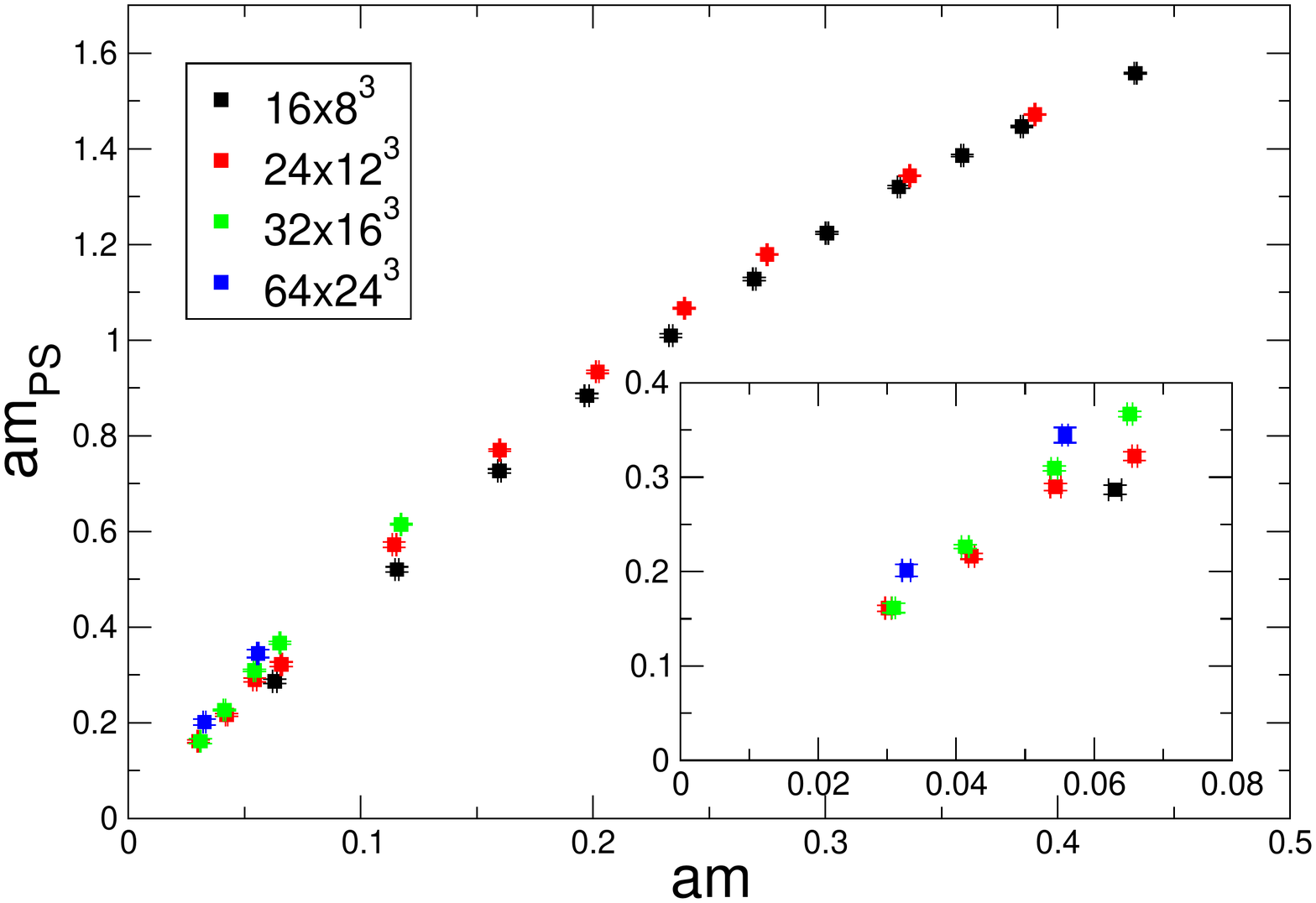}
\label{mpsplot}
}
\qquad
\subfloat[Ratio of $m_{PS}$  to $m$.]{
\includegraphics[scale=0.21]{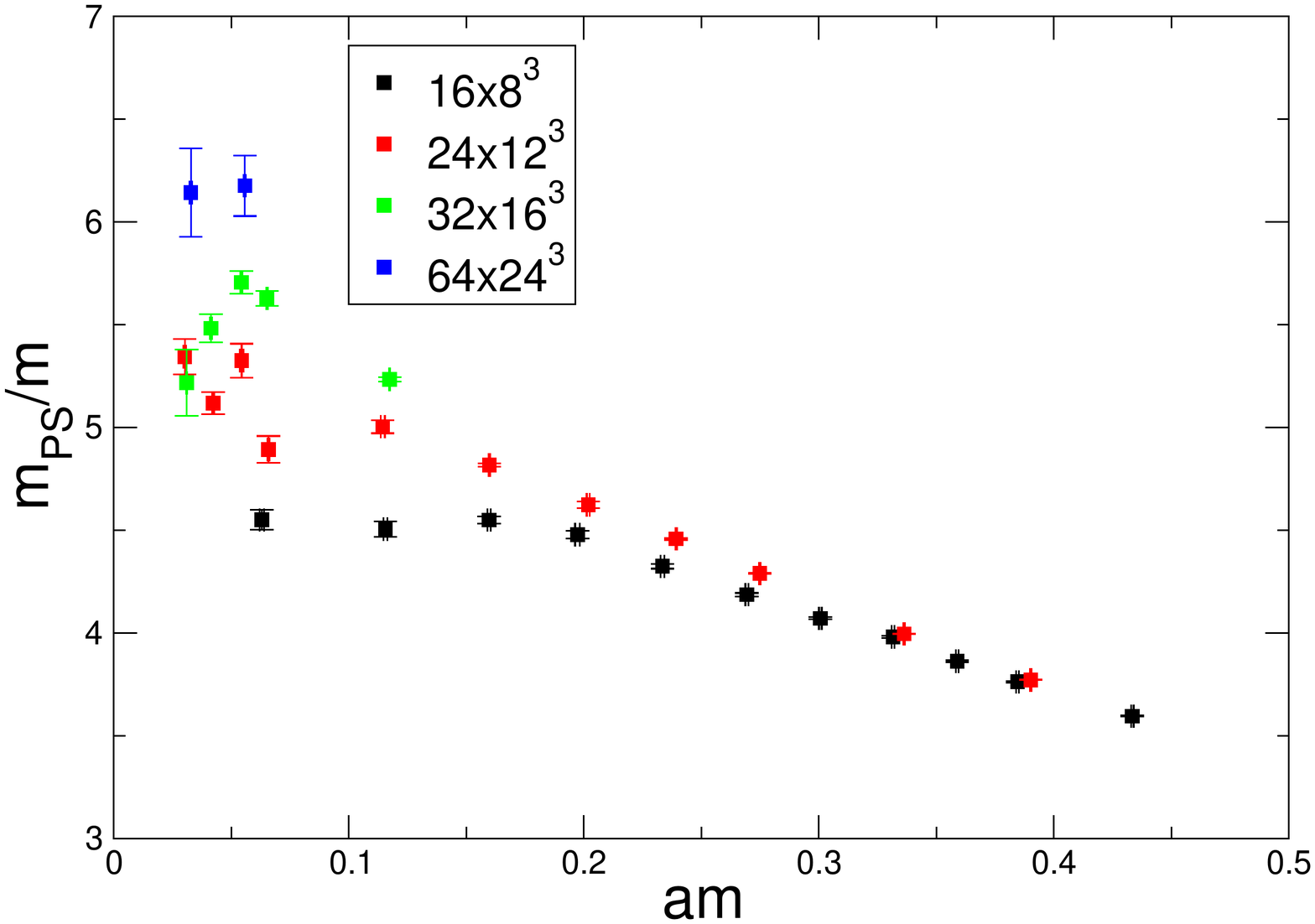}
\label{mps_m_plot}
}
\subfloat[Ratio of $m_{PS}^2$ to $m$.]{
\includegraphics[scale=0.21]{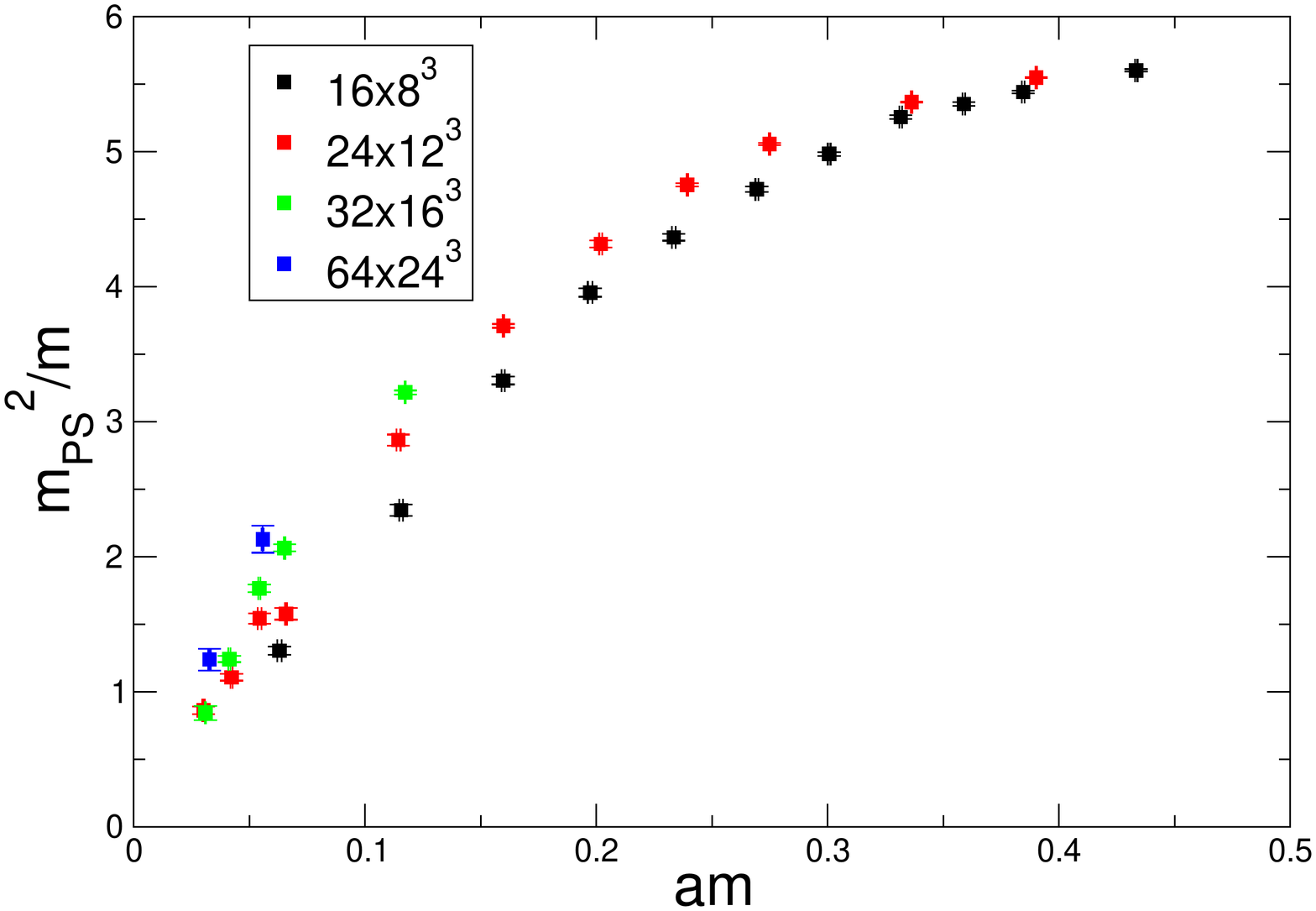}
\label{mps2_m_plot}
}
\qquad
\subfloat[Vector mass]{
\includegraphics[scale=0.21]{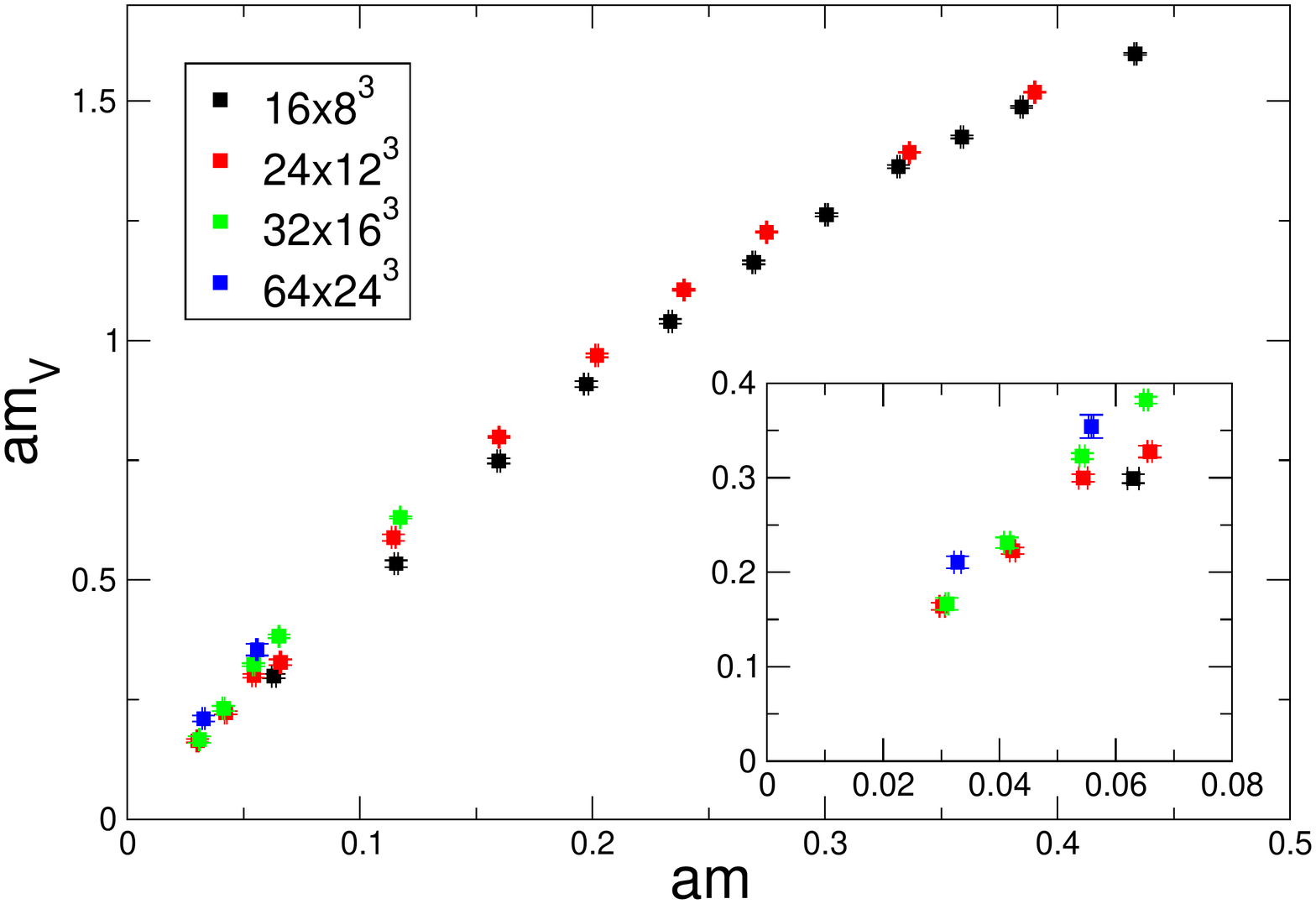}
\label{mvplot}
}
\subfloat[Ratio of vector mass to pseudoscalar mass.]{
\includegraphics[scale=0.21]{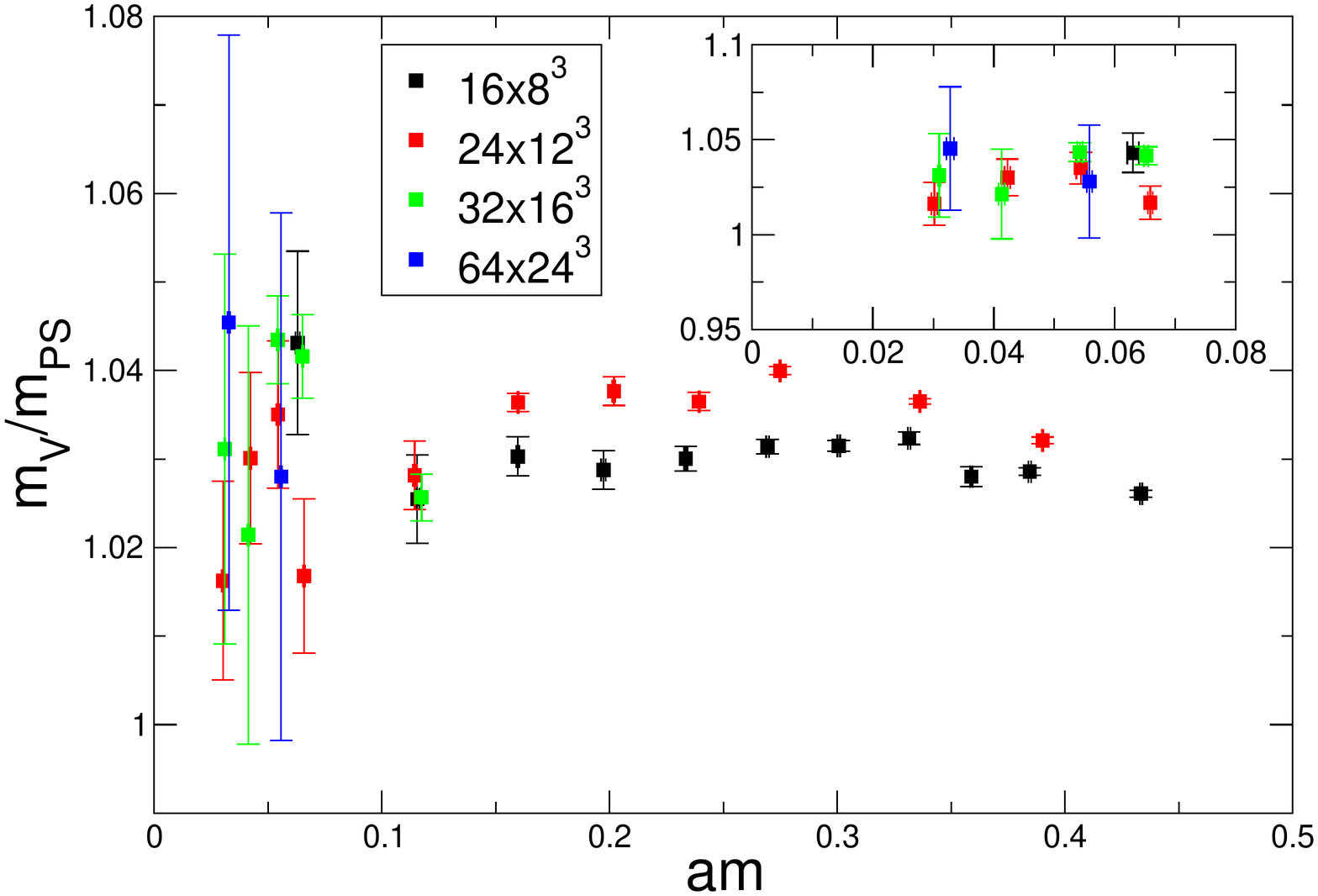}
\label{mvmpsplot}
}
\qquad
\subfloat[Pseudoscalar decay constant.]{
\includegraphics[scale=0.21]{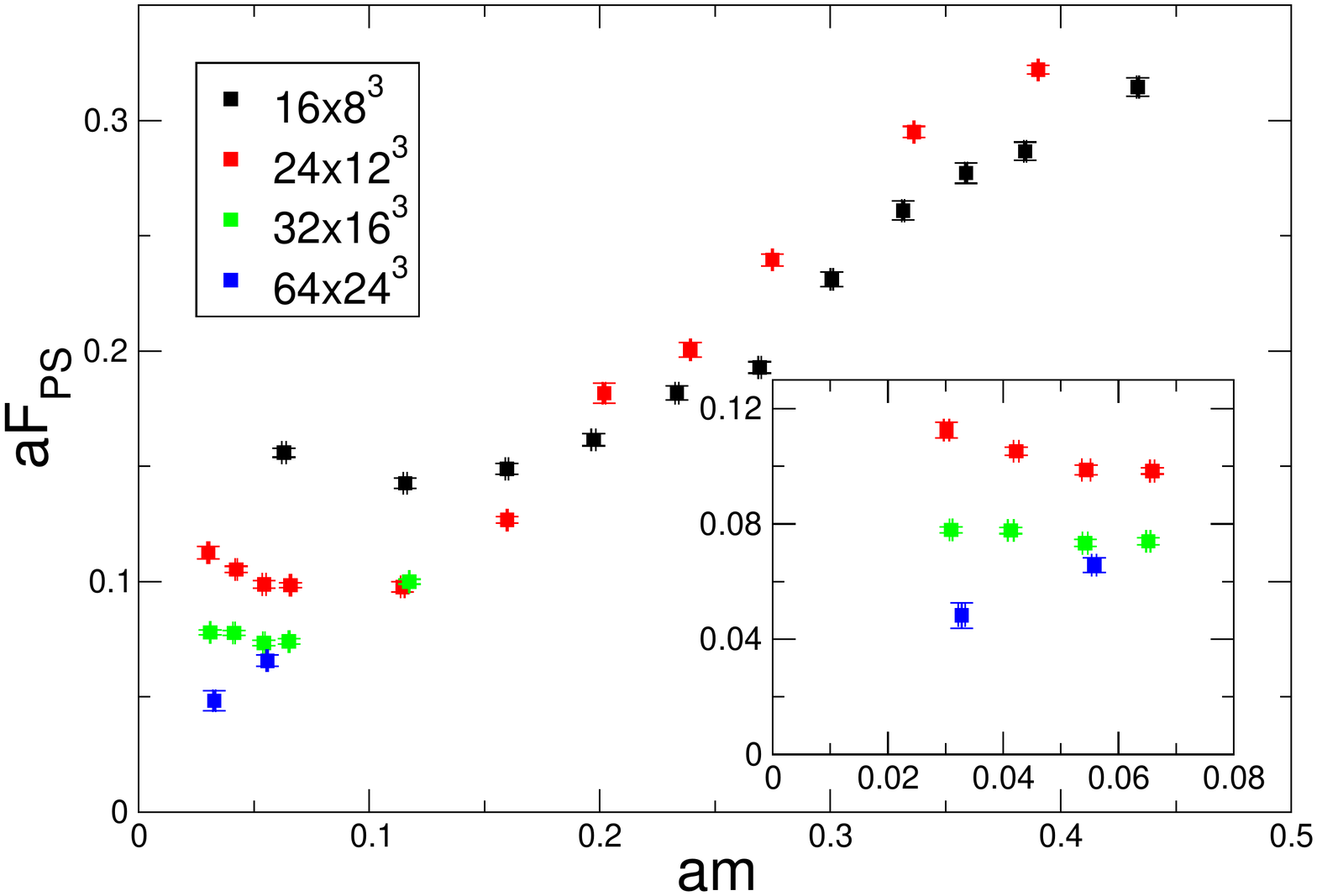}
\label{fpsplot}
}
\subfloat[Chiral condensate as given by the GMOR relation.]{
\includegraphics[scale=0.21]{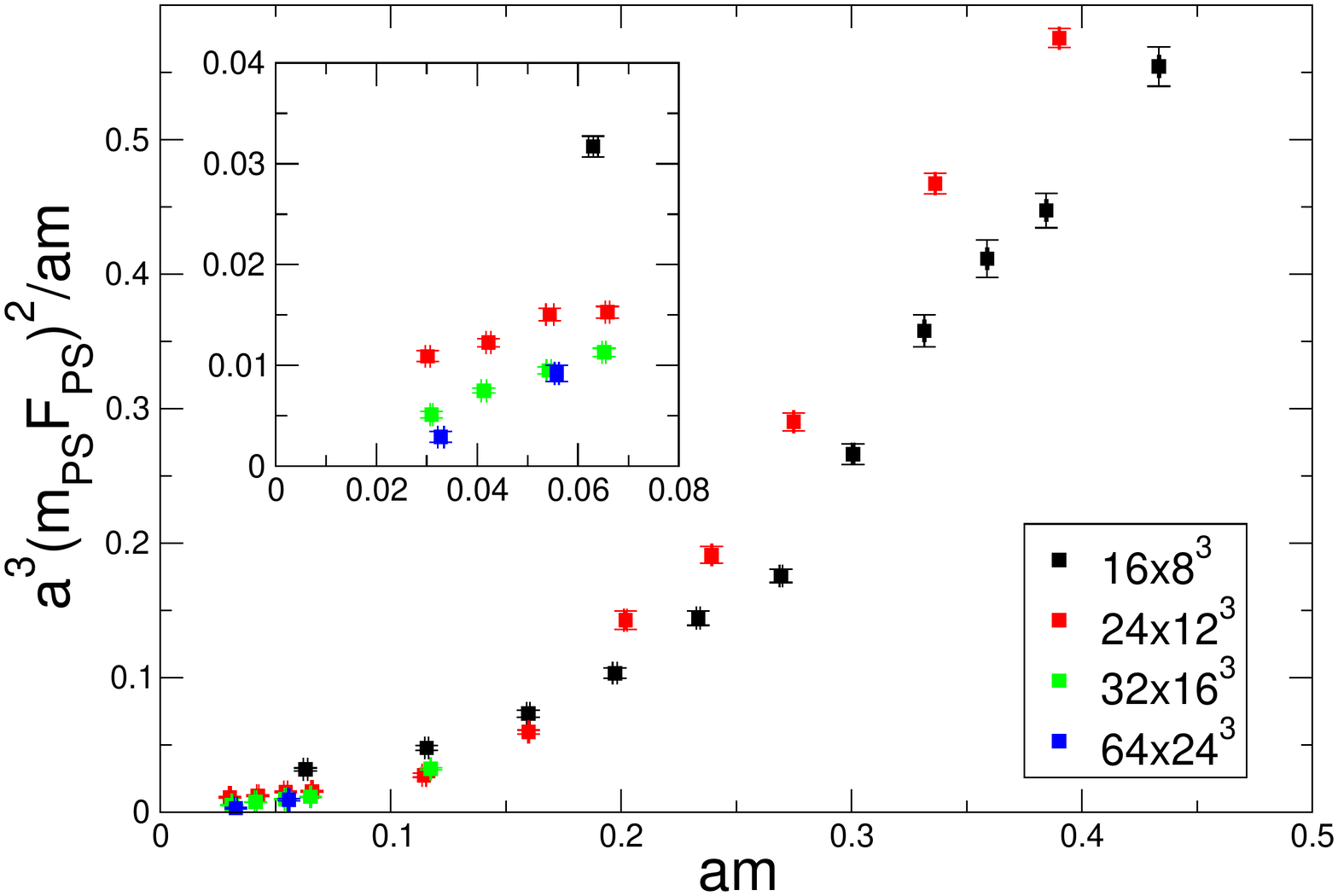}
\label{gmorplot}
}
\caption{Fermionic observables in the chiral region.}
\label{fpsplots}
\end{figure}

Fig. \ref{mpsplot} shows the results obtained for the pseudoscalar mass $m_{PS}$ as a function of the PCAC quark mass $m$. In Figs. \ref{mps_m_plot},
\ref{mps2_m_plot} we can see the ratio of $m_{PS}$ and $m_{PS}^2$, respectively, to $m$. We notice that $\frac{m_{PS}}{m}$ appears to increase in the chiral
limit, while $\frac{m_{PS}^2}{m}$ appears to vanish at zero quark mass. This is at odds with the behaviour expected in a theory described by a chiral
effective field theory (like QCD) in which the pseudoscalar mass is expected to scale as $m_{PS}\sim \sqrt{m}$ in the chiral limit.

In Fig. \ref{mvplot} we see the mass of the vector meson $m_V$ as a function of the quark mass $m$. We can see that it appears to vanish in the
chiral limit, indicating that chiral symmetry is unbroken in this theory. Analysing the ratio of $m_V$ to $m_{PS}$, Fig. \ref{mvmpsplot}, we see
that this does not diverge in the chiral limit, in fact it deviates little from unity, again in conflict with the expectations from a chirally
broken theory where the pseudoscalar is expected to become massless in the chiral limit, with the rest of the mesonic spectrum retaining a finite mass. The
behaviour we observe is more akin to that expected in a mass deformed conformal gauge theory where the property of hyperscaling insists that all masses must
scale to zero with the same critical exponent in the chiral limit.

\subsection{Decay constants}

In Fig. \ref{fpsplot} we show the measured pseudoscalar decay constant as a function of quark mass. Clearly, large finite-volume effects are present,
indicated by the large discrepancies between results on different volumes. 
Using the Gell-Mann--Oakes--Renner relation we can analyse the scaling of the chiral condensate with the quark mass. Although again our data, shown in Fig.
\ref{gmorplot}, suffer from significant finite-size effects, there is no indication of a finite chiral condensate in the chiral limit.

\subsection{Scaling and anomalous dimension}

In Sec. \ref{conformal} we described how the meson masses are expected to scale to zero with the same critical exponent $\rho=\frac{1}{1+\gamma_\ast}$ in the
chiral limit. We have attempted to fit the behaviour of $m_{PS}$ and $m_V$ to such a power law and extract from these fits a value of $\gamma_\ast$. The
$\chi$-squared value of these fits is generally rather poor, except in a few cases. However the results indicate that no particular non-zero value of
$\gamma_\ast$ is preferred by our data, and we see this as consistent with a low value $\gamma_\ast\ll 1$. We have also attempted to determine
$\gamma_\ast$ from the finite-size scaling relations described in Sec. \ref{conformal}. This analysis also indicates a preference for a low $\gamma_\ast$ from
the data.  For details, see \cite{Kerrane:2010up}.

\section{Conclusions}

Our investigation of the systematic effects present in our analysis of mesonic observables in MWT indicates that they are largely under control, and finite
volume effects are small on our larger lattices. We find further supporting evidence for the case that this theory is near-conformal, with indications
that the anomalous dimension at the fixed point in the massless theory is small.
This could be a problem for certain models which depend on a large $\gamma_\ast$ to satisfy current experimental data, however it should be remembered
that MWT must be embedded in a larger theory of extended technicolor in order to construct a model of electro-weak symmetry breaking and as
such $\gamma_\ast$ could be affected by other sectors of the theory, and so our result is nothing like the final answer. 

\section*{Acknowledgements}

The numerical calculations presented in this work have been performed
on the Horseshoe6 cluster at the University of Southern Denmark (SDU)
funded by the Danish Centre for Scientific Computing for the project
``Origin of Mass'' 2009/2010. The author is supported by SUPA, the Scottish Universities Physics Alliance. 
A.R. thanks the Deutsche Forschungsgemeinschaft for financial support. BL is supported by the Royal Society. A.P. was supported by the EC (Research Infrastructure Action in FP7, project \textit{HadronPhysics2}). The development of the code used in this work was partially supported by the EPSRC grant EP/F010303/1.

\bibliographystyle{unsrt}
\bibliography{../Refs/refs}

\end{document}